\documentclass{aa}
\usepackage{natbib}
\usepackage{ucs}
\usepackage[utf8x]{inputenc}
\usepackage{amsfonts}
\usepackage{graphicx}
\usepackage{amssymb}
\usepackage{float}
\usepackage{amsmath}
\usepackage{hyperref}
\usepackage{fontenc}
\date{Submitted}
\begin{document}
\title{Extending the $L_{\mathrm{X}}-T_{\mathrm{}}$ relation from clusters to groups}
\subtitle{Impact of cool core nature, AGN feedback, and selection effects}
\author{V.~Bharadwaj, T.~H.~Reiprich, L.~Lovisari, H.~J.~Eckmiller}
\institute{Argelander-Institut f\"ur Astronomie, Auf dem H\"ugel 71, 53121
  Bonn, Germany}
  \date{Accepted}
\abstract{}{We aim to investigate the bolometric $L_{\mathrm{X}}-T$ relation for galaxy groups, and study the impact of gas cooling, feedback from supermassive black holes, and selection effects on it.}{With a sample of 26 galaxy groups we obtained the best fit $L_{\mathrm{X}}-T$ relation for five different cases depending on the ICM core properties and central AGN radio emission, and determined the slopes, normalisations, intrinsic and statistical scatters for both temperature and luminosity. Simulations were undertaken to correct for selection effects (e.g. Malmquist bias) and the bias corrected relations for groups and clusters were compared.}{The slope of the bias corrected $L_{\mathrm{X}}-T$ relation is marginally steeper but consistent with clusters ($\sim 3$). Groups with a central cooling time less than 1 Gyr (SCC groups) show indications of having the steepest slope and the highest normalisation. For the groups, the bias corrected intrinsic scatter in $L_{\mathrm{X}}$ is larger than the observed scatter for most cases, which is reported here for the first time. Lastly, we see indications that the groups with an extended central radio source have a much steeper slope than those groups which have a CRS with only core emission. Additionally, we also see indications that the more powerful radio AGN are preferentially located in NSCC groups rather than SCC groups.}{}
\titlerunning{$L_{\mathrm{X}}-T_{}$ relation in groups-cool core, feedback, and selection effects}
\authorrunning{V.~Bharadwaj et al.}
\maketitle
\section{Introduction}
A naive starting point when defining groups of galaxies is to simply call them scaled down versions of galaxy clusters. This is not entirely out of context, as the distinction between `groups' and `clusters' is quite loose and no universal definition exists in literature. However, groups and clusters have some notable differences such as a lack of dominance of the intracluster medium (ICM)\footnote{We refrain from the use of the abbreviation IGM to avoid confusion with the intergalactic medium.} over the galactic component (e.g. \citealt{2009ApJ...703..982G}). Moreover, due to the shallower gravitational potential of groups, one would expect processes like AGN heating and galactic winds to leave stronger imprints on the ICM than in clusters. Due to the shape of the cluster mass function (e.g.~\citealt{2008ApJ...688..709T}, \citealt{2009ApJ...692.1033V}), most `clusters' are in fact groups of galaxies, and consequently, most galaxies in the local Universe are present in groups. The upcoming X-ray telescope eROSITA \citep{2010SPIE.7732E..0UP} on the Spektrum-Roentgen-Gamma (SRG) mission promises to detect $10^{5}$ clusters, most of which will be galaxy groups \citep{2012MNRAS.422...44P}. Most of these systems would lack sufficient X-ray counts to be able to measure temperature and mass accurately \citep{2014A&A...567A..65B}, meaning an observable proxy such as luminosity, or external constraints such as weak lensing follow-up~\citep{2012arXiv1209.3114M} would have to be used to constrain these properties. In order to achieve this, the construction of robust, precise, and unbiased scaling relations using existing data gains paramount importance. 

Scaling relations such as $L_{\mathrm{X}}-M_{\mathrm{tot}}$, $L_{\mathrm{X}}-T$ and $M_{\mathrm{tot}}-T$ have been well established for different samples of galaxy clusters (e.g.~\citealt{2001MNRAS.328L..37A, Reiprich-Boehringer:02}, \citealt{2006ApJ...640..691V}; see \citealt{2013SSRv..177..247G} for a review). The validity of cluster scaling relations on the group scale has led to conflicting viewpoints with \cite{ 1996MNRAS.283..690P} suggesting variations for galaxy groups and e.g.~\cite{2009ApJ...693.1142S} refuting any discrepancy between clusters and groups. \cite{ Eckmiller} show that there is a good agreement between clusters and groups for most scaling relations, albeit with increased intrinsic scatter on the group regime-not totally unexpected due to the complicated baryonic physics at play in low mass systems.Recent results by \cite{2014arXiv1409.3845L} however argue that while the \textit{observed} scaling relations for galaxy groups are in agreement with the ones obtained for galaxy clusters, the scaling relation obtained after corrections for selection effects are consistent with a gradual steepening towards the low mass regime.

The $L_{\mathrm{X}}-T_{\mathrm{}}$ relation is one of the most contentious scaling relations. Self-similarity predicts that the bolometric $L_{\mathrm{X}}$ is proportional to $T^{2}$ (e.g.~\citealt{1998ApJ...503..569E}) for clusters, but observations constrain a much steeper slope (e.g.~\citealt{1999MNRAS.305..631A}). Moreover, as compared to other scaling relations, the intrinsic scatter is also much larger (e.g.~\citealt{2009A&A...498..361P}). A major contribution to the scatter in the scaling relation is the cooling gas in the cores of clusters, as shown by e.g.~\cite{2006ApJ...639...64O}. \cite{2011A&A...532A.133M} in particular demonstrate that excising the core regions reduce the scatter in the $L_{\mathrm{X}}-T$ relation by 27\%, additionally stating that cool cores cannot be the sole contributor to the scatter. They also speculate that while ICM cooling dominates on the cluster regime, AGN feedback could have a greater effect on the scaling relation in groups, mainly due to the shallower gravitational potential of these systems.

This paper is the first systematic attempt to determine bolometric $L_{\mathrm{X}}-T$ relations to account for both the presence/absence of a strong cool core and the presence/absence of a central radio loud AGN in galaxy groups. Additionally, we construct a bias corrected bolometric $L_{\mathrm{X}}-T$ relation to objectively compare groups and clusters of galaxies for the first time. The overarching theme of this paper is to understand the impact of non-gravitational processes and selection effects on the $L_{\mathrm{X}}-T$ relation.

A $ \Lambda$CDM cosmology with $ \Omega_{\mathrm{m}} = 0.27$, $\Omega_{\Lambda} = 0.73$ and $ h = 0.71$ where $H_{0} = 100 h$ km/s/Mpc is assumed throughout the paper, unless stated otherwise. All errors are quoted at the 68\% level unless stated otherwise. Log is always base 10 here.

\section{Data and Analysis}
\subsection{Sample and previous work}
We describe briefly the sample and methods which were used for this study. Detailed explanations about the sample and the data reduction can be found in \cite{Eckmiller} and \cite{2014arXiv1402.0868B} respectively.

The sample of galaxy groups was compiled from three X-ray catalogues based on the ROSAT all sky survey by \cite{Eckmiller} in order to test scaling relations on the group regime. Essentially, an upper luminosity cut of $2.55\cdot10^{43}h_{70}^{-2} \mathrm{erg}~\mathrm{s}^{-1}$ and a lower redshift cut of 0.01 was applied to the NORAS catalogue \citep{Boehringer-Voges-Huchra:00}, REFLEX catalogue \citep{ Boehringer-Schuecker-Guzzo:04} and HIFLUGCS \citep{Reiprich-Boehringer:02} clusters, to select a statistically complete sample from which a sub-sample of groups with Chandra data were used for analysis. In the end, 26 objects were used for testing scaling relations.

In a follow-up study, \cite{2014arXiv1402.0868B} investigated the cores of these galaxy groups by determining the temperatures and densities to constrain their cool-core properties such as central cooling time (CCT) and central entropy. Using the CCT as the parameter of distinction, the sample was divided into the strong cool core (SCC with CCT $<$ 1 Gyr), weak cool core (or WCC with 1 Gyr $\leq$ CCT $<$ 7.7 Gyr), and non cool core (or NCC with CCT $\geq$ 7.7 Gyr) classes, where the CCT was determined at 0.4\%$r_{500}$, consistent with the work of \cite{2010A&A...513A..37H} for the HIFLUGCS clusters. The fractions of SCC, WCC and NCC groups were found to be similar to that of clusters. Using radio catalogue data, the presence of central radio sources (CRS) was identified and the radio output, which is a measure of the AGN activity, was determined. Furthermore, with the help of near-infrared data from the 2MASS XSC catalog \citep{2000AJ....119.2498J, 2006AJ....131.1163S}, the brightest cluster galaxy (BCG) was also studied and linked to the ICM cooling and AGN heating, to give a complete picture on the cores of galaxy groups. When the results for the groups were compared to that of clusters, five important differences were identified, e.g.~groups do not follow the trend of clusters to exhibit a higher AGN fraction with decreasing CCT.

\subsection{Temperatures and luminosities}
The data reduction in this work was performed using CIAO 4.4 with CALDB 4.5.0. The \verb+chandra_repro+ task was used to reprocess the raw data set and create a new level 2 event file. The \verb+lc_clean+ routine was used to filter out soft-proton flares. Point sources were detected and excluded using the \verb+wavedetect+ wavelet algorithm. These steps were exactly the same as was implemented in \cite{2014arXiv1402.0868B}. Spectra were extracted in a single annulus centred on the emission weighted centre, excluding the core regions using the radii stated in \cite{Eckmiller} (central region with a cooler temperature component indicated by a central temperature drop in the temperature profile), to get a core-excised temperature and to prevent biasing the temperature estimation to a lower value. 

\begin{figure}
 \includegraphics[angle=270,scale = 0.35]{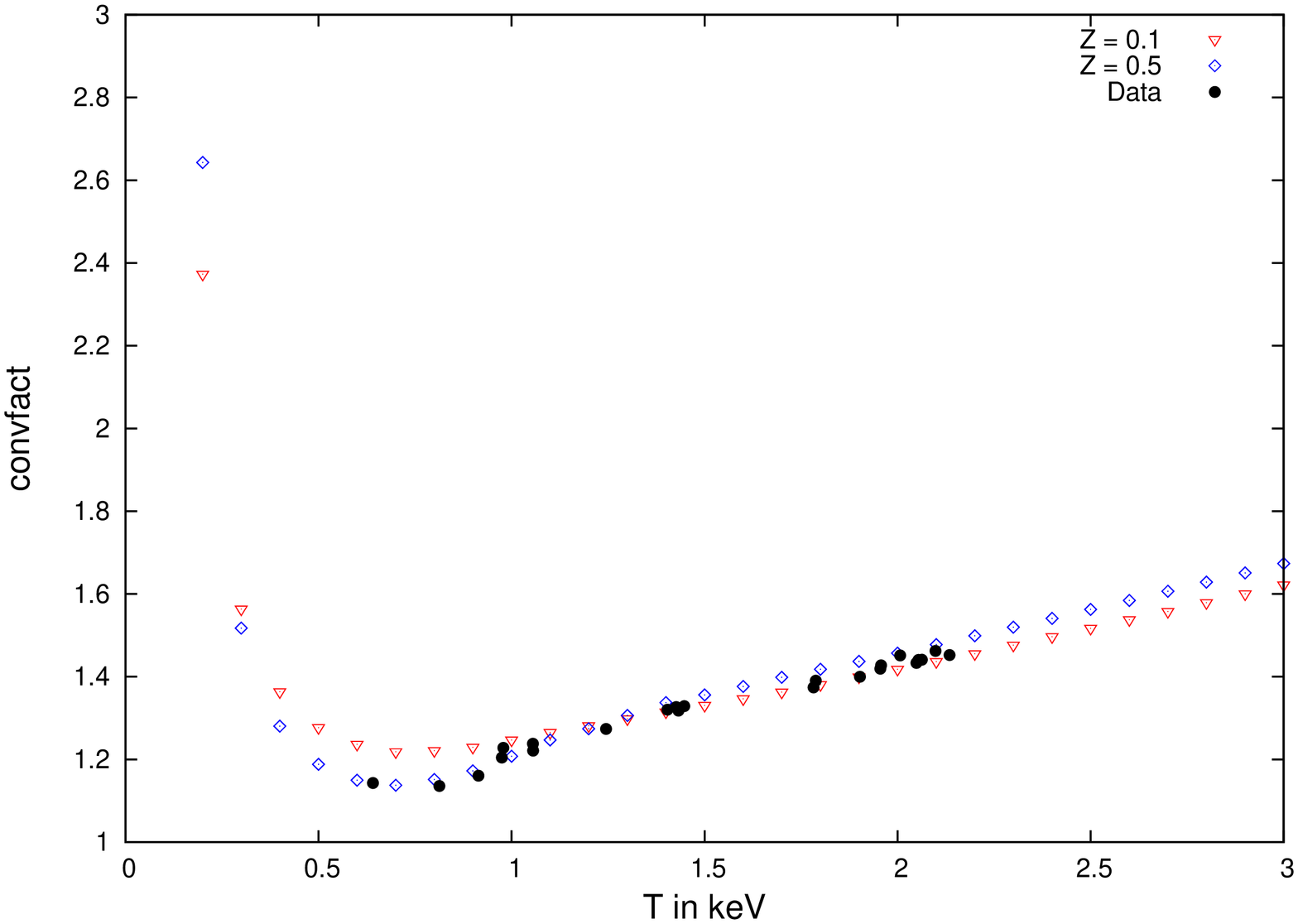}
\caption{Plot of conversion factor between ROSAT and bolometric luminosities as a function of temperature. Red triangles are for metallicity of 0.1, blue diamonds are for metallicity of 0.5, black points represent the actual conversion factors used for the sample.}
\label{confact}
\end{figure}

The background treatment was performed like in \cite{2014arXiv1402.0868B}. The particle background was estimated using the stowed events files distributed within the CALDB. For the astrophysical background, we performed a simultaneous spectral fit to the Chandra data and the ROSAT all sky survey data (provided by Snowden's webtool\footnote{\url{http://heasarc.gsfc.nasa.gov/cgi-bin/Tools/xraybg}}). The background components were an absorbed power law with a spectral index of 1.41 for unresolved AGN, an absorbed APEC model for Galactic halo emission, and an unabsorbed APEC model for Local Hot Bubble emission \citep{1998ApJ...493..715S}. The RASS data were taken from an annulus far away from the group centre, where no group emission would be present. The group emission was modelled with an absorbed APEC model, with the temperature and abundance free to vary. All absorption components were linked and modelled with the \textit{phabs} model, and $N_{\mathrm{H}}$ values were taken from a webtool\footnote{\url{http://www.swift.ac.uk/analysis/nhtot/index.php}} which follows the method described in \cite{2013MNRAS.431..394W}. In some cases, these were found to be too low, and they were left thawed in the spectral fit. This has the effect of lowering the temperatures, with the largest change being in the order of 6\%. The \cite{1989GeCoA..53..197A} abundance table and AtomDB 2.0.2 was used throughout. Since we wished to study the bolometric $L_{\mathrm{X}}-T$ relation, we had to convert the quoted ROSAT (0.1-2.4 keV) luminosities in \cite{Eckmiller} to the bolometric band (0.01-40 keV), for which we used the program \textit{Xspec}. Using an APEC model frozen to the best fit temperature, abundance and redshift for the groups, we calculated the luminosities in the ROSAT band and the bolometric band. The ratio of the ROSAT and bolometric $L_{\mathrm{X}}$ gives us the conversion factor to transform between the two luminosities. This conversion factor largely depends on the temperature and slightly on the abundance of the ICM (Fig.~\ref{confact}). Given that most groups from the eROSITA all sky survey would likely lack sufficient counts to resolve the core, we did not use core-excised luminosities. The temperatures, luminosities and $N_{\mathrm{H}}$ values are listed in Table~\ref{TLtab}.

\begin{table*}
 \centering
 \caption{Temperatures, bolometric luminosities and $N_{\mathrm{H}}$ for the galaxy groups. Starred entries represent fitted values for $N_{\mathrm{H}}$.}\label{TLtab}
 \begin{tabular}{|c|c|c|c|}
  \hline & & & \\
  Group name & $T$ keV& $L_{\mathrm{X}}~10^{43}$ erg/s & $N_{\mathrm{H}} 10^{21}~\mathrm{cm}^{-2}$ \\\hline & & & \\
  A0160 &$1.90^{+0.09}_{-0.09}$ &$2.69^{+0.38}_{-0.38} $ &$0.516 $ \\ & & & \\
  A1177 &$1.79^{+0.07}_{-0.07}$ &$1.22^{+0.17}_{-0.17} $ &$0.117 $ \\ & & & \\
  ESO55 &$2.10^{+0.05}_{-0.07}$ &$2.02^{+0.24}_{-0.24} $ &$1.02^{*} $ \\ & & & \\
  HCG62 &$1.43^{+0.04}_{-0.05}$ &$0.399^{+0.053}_{-0.053} $ &$1.03^{*} $ \\ & & & \\
  HCG97 &$0.98^{+0.02}_{-0.02}$ &$0.358^{+0.139}_{-0.139} $ &$1.56^{*} $ \\ & & & \\
  IC1262 &$1.95^{+0.04}_{-0.05}$ &$3.42^{+0.17}_{-0.17} $ &$0.416^{*} $ \\ & & & \\
  IC1633 &$3.58^{+0.14}_{-0.14}$ &$3.10^{+0.25}_{-0.25} $ &$0.200 $ \\ & & & \\
  MKW4 &$2.01^{+0.03}_{-0.03}$ &$2.86^{+0.05}_{-0.05} $ &$0.534^{*} $ \\ & & & \\
  MKW8 &$3.25^{+0.08}_{-0.08}$ &$6.92^{+0.58}_{-0.58} $ &$0.269 $ \\& & & \\
  NGC326 &$2.06^{+0.08}_{-0.09}$ &$3.22^{+0.42}_{-0.42} $ &$1.07^{*} $ \\& & & \\
  NGC507 &$1.45^{+0.02}_{-0.02}$ &$1.67^{+0.02}_{-0.02} $ &$0.638 $ \\& & & \\
  NGC533 &$1.43^{+0.04}_{-0.04}$ &$0.413^{+0.066}_{-0.066} $ &$0.331 $ \\& & & \\
  NGC777 &$0.81^{+0.02}_{-0.02}$ &$0.232^{+0.032}_{-0.032} $ &$0.570$ \\& & & \\
  NGC1132 &$1.25^{+0.01}_{-0.01}$ &$0.929^{+0.150}_{-0.150} $ &$0.556 $ \\& & & \\
  NGC1550 &$1.40^{+0.01}_{-0.01}$ &$2.00^{+0.11}_{-0.11} $ &$1.62 $ \\& & & \\
  NGC4325 &$1.06^{+0.01}_{-0.01}$ &$1.25^{+0.10}_{-0.10} $ &$1.41^{*} $ \\& & & \\
  NGC4936 &$0.97^{+0.01}_{-0.02}$ &$0.293^{+0.040}_{-0.040} $ &$0.882 $ \\& & & \\
  NGC5129 &$1.05^{+0.01}_{-0.01}$&$0.445^{+0.076}_{-0.076} $ &$0.186 $ \\& & & \\
  NGC5419 &$1.96^{+0.12}_{-0.15}$ &$0.444^{+0.064}_{-0.064} $ &$1.44^{*} $ \\& & & \\
  NGC6269 &$2.05^{+0.10}_{-0.12}$ &$2.66^{+0.22}_{-0.22} $ &$0.849^{*} $ \\& & & \\
  NGC6338 &$2.13^{+0.10}_{-0.05}$ &$3.63^{+0.56}_{-0.56} $ &$0.400^{*} $ \\& & & \\
  NGC6482 &$0.64^{+0.01}_{-0.01}$&$0.111^{+0.012}_{-0.012} $ &$1.04 $ \\& & & \\
  RXCJ1022 &$2.05^{+0.10}_{-0.12}$ &$2.62^{+0.74}_{-0.74} $ &$0.653^{*} $ \\& & & \\
  RXCJ2214 &$1.43^{+0.06}_{-0.10}$ &$0.602^{+0.109}_{-0.109} $ &$0.605 $ \\& & & \\
  S0463 &$1.78^{+0.29}_{-0.22}$ &$1.98^{+0.33}_{-0.33} $ &$0.084 $ \\& & & \\
  SS2B &$0.91^{+0.01}_{-0.01}$ &$0.666^{+0.045}_{-0.045} $ &$1.66^{*}$ \\& & & \\ \hline
  
 \end{tabular}

\end{table*}

For getting the slope and normalisation of the best fit scaling relation, we used the BCES (Y$\lvert$X) code by \cite{ 1996ApJ...470..706A}. The fits were performed in log space for the fitting function:
\begin{equation}
\left ( \frac{L_{\mathrm{X}}}{0.5\times 10^{44}~\mathrm{erg~s}^{-1}} \right ) = c\times\left ( \frac{T}{3~\mathrm{keV}} \right )^{m}
\end{equation}

We also calculated statistical and intrinsic scatters with the following formulae\footnote{In log space, errors are expressed as $\Delta~\mathrm{log}~x = \mathrm{log}(e)~(x^{+}-x^{-})/(2x)$ where $x^{+}$ and $x^{-}$ are the upper and lower boundary of the error range of the quantity $x$}:

\begin{eqnarray}
\sigma^{T}_{\mathrm{stat}} = \left \langle \log(\mathrm{e})\cdot \Delta T/T \right \rangle \\
\sigma^{L_{\mathrm{X}}}_{\mathrm{stat}} = \left \langle \log(\mathrm{e})\cdot \Delta L_{\mathrm{X}}/L_{\mathrm{X}} \right \rangle \\
\sigma^{T}_{\mathrm{tot}} = \left \langle (\log T-(\log L_{\mathrm{X}}-\log c)/m)^{2} \right \rangle^{1/2} \\
\sigma^{L_{\mathrm{X}}}_{\mathrm{tot}} = \left \langle (\log L_{\mathrm{X}}-(m\cdot \log T + \log c))^{2} \right \rangle^{1/2} \\
\sigma^{T}_{\mathrm{int}} = \left ( (\sigma^{T}_{\mathrm{tot}})^{2} - (\sigma^{T}_{\mathrm{stat}})^{2} - m^{-2}\cdot(\sigma^{L_{\mathrm{X}}}_{\mathrm{stat}})^{2} \right )^{1/2} \\
\sigma^{L_{\mathrm{X}}}_{\mathrm{int}} = \left ( (\sigma^{L_{\mathrm{X}}}_{\mathrm{tot}})^{2} - (\sigma^{L_{\mathrm{X}}}_{\mathrm{stat}})^{2}-m^{2}\cdot(\sigma^{T}_{\mathrm{stat}})^{2} \right )^{1/2}
\end{eqnarray}

The fits were performed for five different cases, which are presented in Table~\ref{sumtable}. To test the effects of ICM cooling on the relation, we sub-divided the sample into the SCC and non-strong cool core (NSCC; CCT $\geq~1$ Gyr) cases. To test the effects of AGN feedback, we classified the systems as those with and without a central radio source (CRS and NCRS respectively), a CRS representative of the presence of a radio loud AGN. 

\begin{table*}[]
  \centering
  \caption{Observed bolometric $L_{\mathrm{X}}-T_{\mathrm{}}$ relation.}\label{sumtable} 
  \begin{tabular}{|c|c|c|c|c|c|c|}
    \hline
    Category & slope & normalisation 
    & $\sigma_{{\mathrm{int},~L_{\mathrm{X}}}}$   & $\sigma_{\mathrm{stat},~L_{\mathrm{X}}}$ 
    & $\sigma_{\mathrm{int},~T_{\mathrm{vir}}}$ & $\sigma_{\mathrm{stat},~T_{\mathrm{vir}}}$\\
    \hline \hline   
    All Groups & 2.17$\pm$0.26 & -0.01$\pm$0.09&0.237&0.056&0.109&0.015\\
    All HIFLUGCS clusters& 2.97$\pm$0.20&0.42$\pm$0.04&0.264&0.010&0.089&0.013\\
   \hline 
    SCC Groups  &2.56$\pm$0.22&0.17$\pm$0.10&0.230&0.054&0.090&0.0099\\
    SCC Clusters&3.46$\pm$0.20&0.51$\pm$0.05&0.234&0.0092&0.068&0.0093\\
    \hline
    NSCC Groups  &2.00$\pm$0.39&-0.09$\pm$0.11&0.226&0.059&0.113&0.019\\
    NSCC Clusters &2.76$\pm$0.29&0.40$\pm$0.05&0.240&0.011&0.087&0.015\\
    \hline
    CRS Groups  &2.14$\pm$0.31&-0.03$\pm$0.11&0.253&0.049&0.118&0.015\\
    CRS Clusters &3.31$\pm$0.20&0.45$\pm$0.04&0.237&0.011&0.072&0.012 \\
    \hline
    NCRS Groups  &2.29$\pm$0.31&0.04$\pm$0.09&0.169&0.081&0.074&0.013\\
    NCRS Clusters &2.40$\pm$0.37&0.41$\pm$0.11&0.227&0.0082&0.095&0.014\\
    \hline
\end{tabular}
\end{table*}

\subsection{Bias correction}

Observed scaling relations can be affected by various selection biases, chief among which is the Malmquist bias. The Malmquist bias is the preferential detection of brighter objects for e.g. a given temperature, in a flux-limited sample due to intrinsic scatter. It is an effect which has been shown to affect scaling relations, namely resulting in higher observed normalisations as compared to the actual normalisations in scaling relations for clusters (e.g.~\citealt{2002A&A...383..773I}, \citealt{2006ApJ...648..956S}, \citealt{2009A&A...498..361P}, \citealt{2010MNRAS.406.1773M}, \citealt{2014arXiv1409.3845L}). In this sample, we have applied an additional luminosity cut which could further contribute to the bias. Thus, in order to determine the `true' relation, one has to correct for these biases. To do this, we undertook simulations, the procedure of which we describe here:

For the simulations, we randomly generated samples of groups with temperatures between 0.6 and 3.6 keV. These groups were assigned luminosity distances such that the number of objects scaled as $D_{\mathrm{L}}^{3}$, between redshifts of 0.01 and 0.2, and temperatures corresponding to the X-ray temperature function (the XTF, \citealt{1998ApJ...504...27M}, given by $\mathrm{d}N/\mathrm{d}V \sim T^{-3.2}$).  Using different combinations of slopes, normalisations and scatters, we assigned luminosities to these temperatures. The intrinsic scatter about $L_{\mathrm{X}}$ is log-normal and was introduced as such. Measurement errors were included for both $L_{\mathrm{X}}$ and $T$ which corresponded to the observed statistical scatter. For each set of parameters, we generated 100 samples and applied a flux cut of $2.5\times 10^{-12}~\mathrm{erg~s}^{-1}~\mathrm{cm}^{-2}$ and an upper luminosity cut of $2.55\times 10^{43}h_{70}^{-2}~\mathrm{erg}~\mathrm{s}^{-1}$ (both in the ROSAT band) to simulate the selection criteria, with each sample containing between 20-30 objects, similar to the actual group sample used in the study. To determine the best set of parameters, we defined a chi-squared:

\begin{equation}
\chi^{2} \equiv 
 \left ( \frac{m_{\mathrm{obs}}-m_{\mathrm{out}}}{\Delta m_{\mathrm{obs}}} \right )^{2} + \left ( \frac{c_{\mathrm{obs}}-c_{\mathrm{out}}}{\Delta c_{\mathrm{obs}}} \right )^{2} + \left ( \frac{\sigma_{\mathrm{obs}}-\sigma_{\mathrm{out}}}{\Delta \sigma_{\mathrm{obs}}} \right )^2
\end{equation}

\begin{figure}
 \includegraphics[scale=0.25]{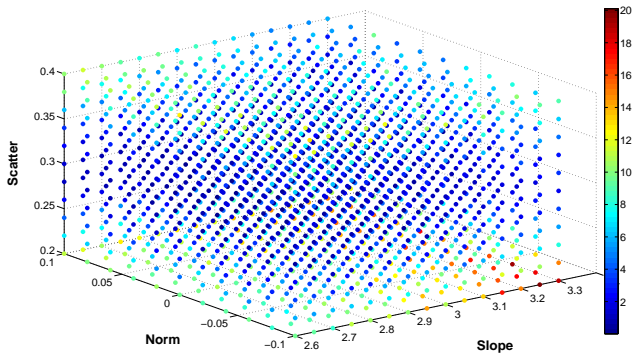}
\caption{Slopes, normalisations, and intrinsic scatter for the `All groups' case. The plot is colour-coded to represent the value of chi-squared.}\label{params}
\end{figure}

Here $m$ is slope, $c$ is normalisation, $\sigma$ is the intrinsic scatter in luminosity. Obs is the observed parameters, out is the output given by the simulation. The deltas represent the measured errors, from the BCES fit. For the scatter, we assumed that $\frac{\Delta \sigma_{\mathrm{obs}}}{\sigma_{\mathrm{obs}}} = 10\% $. Changing this value from 5 to 20\% does not alter the results significantly. An example of this simulation is shown in Fig.~\ref{params}.

Scaling relations for different cases, e.g. SCC and NSCC groups, may differ, therefore selection effects may differ for the different cases as well, making it imperative to determine bias corrections for the individual sub-cases. To do this, fresh simulations were run with the fractions of SCC/NSCC objects, and CRS/NCRS objects fixed to the observed values noted in \cite{2014arXiv1402.0868B}. We note that small changes in the SCC/NSCC fractions and the CRS/NCRS fractions do not drastically affect the determination of the bias corrected slopes, normalisations and scatters for the group sample.

Note that, despite our concerted efforts to correct for the selection biases, due to the incomplete nature of this sample compiled from the Chandra archives, we do not rule out the possibility of a potential `archival bias' which could have a bearing on the results and which we cannot correct for.

\subsection{Cluster comparison sample}
To compare our results for groups to galaxy clusters, we decided to use the HIFLUGCS galaxy clusters \citep{Reiprich-Boehringer:02}, a flux-limited sample with ROSAT (0.1-2.4 keV) flux $\geq 2\times 10^{-11}~\mathrm{erg~s}^{-1}~\mathrm{cm}^{-2}$. For this study, we took the bolometric X-ray luminosities and virial temperatures quoted in \cite{2011A&A...532A.133M}. Since the temperatures were determined with CALDB version 3 (3.2.1) versus version 4 for the galaxy groups, we converted the quoted temperatures using the scaling relation quoted in \cite{2011A&A...532A.133M}, namely:
\begin{equation}
 T_{\mathrm{4.1.1}} = 0.875*T_{\mathrm{3.2.1}} + 0.251
\end{equation}
Using this scaling relation has the effect of lowering the temperature, thereby raising the normalisation and steepening the observed cluster $L_{\mathrm{X}}-T$ relation. The change in slope and normalisation however, are within the errors to that obtained with the older temperatures. Since the groups and the cluster samples have different selection criteria, bias corrections were performed for clusters as well, using the above flux limit in order to compare results accurately. For the sub-samples, we took the SCC/NSCC fraction and the CRS/NCRS fraction for the clusters from \cite{2010A&A...513A..37H} and \cite{2009A&A...501..835M} respectively.   

\begin{figure}
 \centering
 \includegraphics[scale = 0.35,angle = 270]{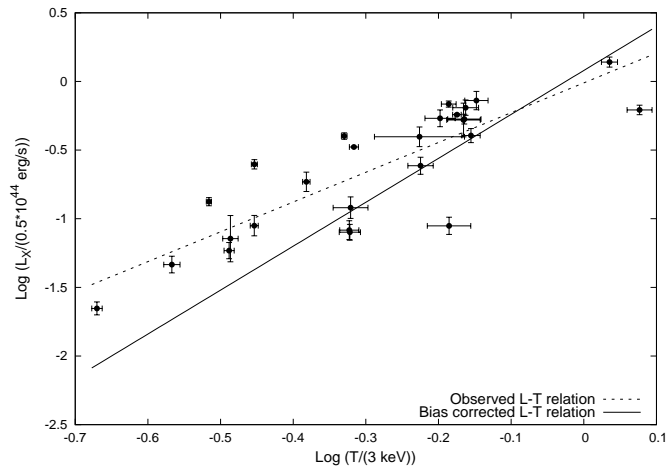}
\caption{Observed and bias corrected $L_{\mathrm{X}}-T$ relation for galaxy groups. The dotted line is the observed scaling relation and the solid line is the bias corrected one.}\label{GLxTobscorr}
\end{figure}

\section{Results and discussion}
\subsection{Observed, bias uncorrected $L_{\mathrm{X}}-T$ relation}

The fit results for the observed data sets (both groups and clusters) are shown in Table~\ref{sumtable}. For all groups, the best fit results are shown in Fig.~\ref{GLxTobscorr}. 

The very first observation that one makes on comparing the observed $L_{\mathrm{X}}-T$ relation of groups to clusters is the flattening of the slope when one enters the group regime. Additionally, we observe that for most cases, the slope of the observed $L_{\mathrm{X}}-T$ relation is consistent with the self-similar value of 2 expected for clusters within the errors. As we demonstrate in Sect.~\ref{biassect}, this is due to the selection criteria applied to construct this sample, and is not a property of the underlying group population. Additionally, the observed $L_{\mathrm{X}}-T$ relation for the SCC groups shows indications of being higher in normalisation (by a factor of $\sim 2$) and having a steeper slope as compared to the NSCC groups, and for groups with a CRS, there is an indication that the normalisation is lower than for those groups without a CRS. 

The statistical scatters in the temperature for both groups and clusters are in good agreement, but for the luminosity, we see an increase by around a factor of 5 as we go from the cluster to the group regime. This is expected, as groups, being low surface brightness systems have a much larger error on their luminosity than clusters. Additionally, the luminosities for the galaxy groups come from the ROSAT all sky-survey (RASS) data with very short exposure times, versus that of the clusters which come from ROSAT position sensitive proportional counter (PSPC) pointed observations with much larger exposure times, leading to a much more precise determination of the cluster luminosities.

\subsection{Bias corrected $L_{\mathrm{X}}-T$ relation}\label{biassect}

We tabulate the bias corrected $L_{\mathrm{X}}-T$ relation for both groups and clusters in Table~\ref{Lbolall}. As pointed out before, this is the first attempt to present a bias corrected bolometric $L_{\mathrm{X}}-T$ relation for groups including their cool-core and central AGN properties, making comparisons between groups and clusters a lot easier than before. 

\begin{table*}
 \centering
\caption{Bias corrected, bolometric $L_{\mathrm{X}}-T$ relation for groups and clusters. Errors are the measurement errors from observations.}\label{Lbolall}
\begin{tabular}{|c|c|c|c|c|} 
 \hline 
Sample & Category & Slope, $(m)$ & Normalisation, $(c)$ & $\sigma_{{\mathrm{int},~L_{\mathrm{X}}}}$ 
    \\ \hline \hline
Groups & ALL &3.20$\pm$0.26&0.08$\pm$0.09&0.32  \\
HIFLUGCS & ALL &2.79$\pm$0.20&0.21$\pm$0.04&0.27  \\ \hline
Groups & SCC &3.60$\pm$0.22&0.30$\pm$0.10&0.30 \\
HIFLUGCS & SCC &3.45$\pm$0.20&0.34$\pm$0.05&0.26  \\ \hline
Groups & NSCC &2.52$\pm$0.39&-0.17$\pm$0.11&0.30   \\
HIFLUGCS & NSCC &2.68$\pm$0.29&0.22$\pm$0.05&0.26   \\ \hline
Groups & CRS &3.60$\pm$0.31&0.16$\pm$0.11&0.36\\
HIFLUGCS & CRS &3.20$\pm$0.20&0.26$\pm$0.04&0.26   \\ \hline
Groups & NCRS &3.20$\pm$0.31&0.09$\pm$0.09&0.30\\
HIFLUGCS & NCRS &2.40$\pm$0.37&0.10$\pm$0.11&0.30 \\ \hline
\end{tabular}
\end{table*}

\begin{figure}
 \centering
\includegraphics[scale=0.35,angle=270]{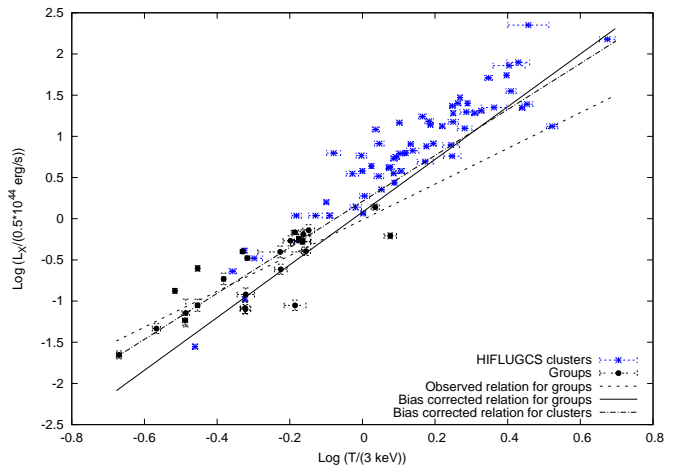}
\caption{Observed and bias corrected $L_{\mathrm{X}}-T$ relations. Clusters are represented by blue points and groups are represented by black points. Dotted line is observed relation for groups, solid line is bias corrected relation for groups, dot-dash line is bias corrected relation for clusters.}\label{trueLT}
\end{figure}

Our very first observation is that correcting for selection effects has a significant impact on the $L_{\mathrm{X}}-T$ relation for the galaxy groups. Over most of the temperature range covered by the group sample, the bias corrected relation results in a lower $L_{\mathrm{X}}$ for a given $T$, and the bias corrected slope steepens significantly from 2.17 to 3.20 (Fig.~\ref{GLxTobscorr}). With the corrections in place, the value of our slope agrees well with previous bolometric $L_{\mathrm{X}}-T$ slopes for galaxy groups observed for incomplete samples (e.g.~\citealt{2004MNRAS.350.1511O}), but is much lower than reported slopes of $\sim 5$ (\citealt{2000MNRAS.315..356H, 2000ApJ...538...65X} at $>~5\sigma$ significance). The value of the corrected slope shows indications of a steepening than that of the corrected slope for clusters, but the two are consistent within the errors (Fig.~\ref{trueLT}, Table~\ref{Lbolall}). Qualitatively, our results show the same trend as reported recently by \cite{2014arXiv1409.3845L}.

\begin{figure}
 \centering
\includegraphics[scale=0.35,angle=270]{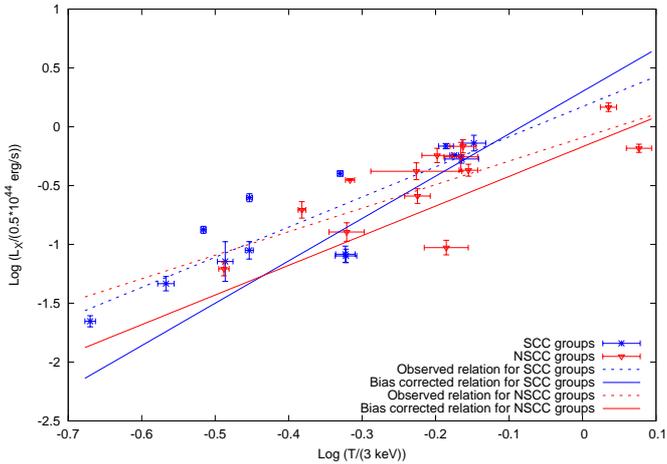}
\caption{Observed and true $L_{\mathrm{X}}-T$ relation for SCC and NSCC groups. Blue dotted line represents observed relation for SCC groups, blue solid line represents true relation for SCC groups, red dotted line represents observed relation for NSCC groups, red solid line represents true relation for NSCC groups.}\label{SCCLT}
\end{figure}

Sub-classifying the sample yields more interesting features. The $L_{\mathrm{X}}-T$ relation for the SCC and CRS groups is by far the steepest with the SCC groups having the highest normalisation for all the sub-samples at 3 keV. The slope of the bias corrected $L_{\mathrm{X}}-T$ relation for the SCC groups is steeper than the NSCC one by $~43\%$ and higher in normalisation by a factor of $~\sim 3$ at 3 keV (Fig.~\ref{SCCLT}). The normalisations for the CRS groups and NCRS groups are in good agreement (within $17\%$, ignoring the errors), but there are indications of marginal steepening of the slopes for the CRS systems, albeit the large error bars make it hard to confirm. The slopes of most of the sub-samples for the groups and clusters are once again consistent within the errors also for the corrected relations. These numbers suggest a rather complicated scenario on the group regime and we discuss this in Sect.~\ref{comppic}.

Our attempt to identify the bias-corrected scatter vs. the observed scatter for $L_{\mathrm{X}}$ to our best knowledge is a first for the $L_{\mathrm{X}}-T$ scaling relation. Interestingly, while the observed and the bias corrected scatter for the complete cluster sample agree within 10\%, the bias corrected scatter for groups is higher by 35\% as compared to the observed one. This statement also holds true qualitatively for most of the sub-samples that we fit here. The reason behind this could simply be due to the applied upper luminosity cut to select the group sample. The corrected intrinsic scatter increases by 19\% as one moves down from the cluster to the group regime, with the largest change in intrinsic scatter between groups and clusters observed for the CRS case ($\sim 38\%$). We have demonstrated for the first time, that the bias-corrected intrinsic scatter in $L_{\mathrm{X}}$ seems to go up from the cluster to the group regime. This would indicate a stronger impact of non-gravitational processes on the group regime than the cluster regime, as one would expect. Interestingly, this conclusion of ours is somewhat different to \cite{2014arXiv1409.3845L} who conclude that the scatter decreases as one goes from the cluster to the group regime. We point out that their conclusion is based on the \textit{observed} scatter as they do not explicitly try to obtain a bias-corrected scatter as we have done in this study. Nevertheless, the small sample sizes in both studies, the choice of X-ray luminosities (ROSAT band vs. bolometric band), potential archival bias in our study, and different techniques to perform the bias correction could all have a potential impact on the determination of the `true' scatter.

Our simulations clearly demonstrate that selection criteria employing a flux and luminosity cut have a stronger impact on the $L_{\mathrm{X}}-T$ relation, particularly on the slope, than those with just a simple flux cut, as in e.g. the HIFLUGCS cluster sample. Additionally, we have also demonstrated that the bias corrections for sub-samples, especially SCC and NSCC sub-samples are not identical and have to be determined individually. 

\subsection{A complete picture of the $L_{\mathrm{X}}-T$ relation}\label{comppic} 
When one wishes to interpret observational results for scaling relations, a good starting point is to look at existing results from simulations. In recent times, the introduction of sophisticated code which accounts for non-gravitational processes such as cooling and feedback have made simulations much more accurate than before. It is all but clear that only simulation recipes with some form of AGN feedback can mitigate excessive gas cooling and high star formation rates which are inconsistent with observations (e.g.~\citealt{2004MNRAS.348.1078B}). Recent results by e.g.~\cite{2008ApJ...687L..53P} and \cite{2010MNRAS.406..822M} argue that a $L_{\mathrm{X}}-T$ relation consistent with observations is obtained for groups only with AGN feedback. This simplified picture would seem to solve all problems, but observations seem to point to a more complex scenario.

Firstly, interpreting the high normalisations of the SCC groups is relatively straightforward; these are systems with the most centrally dense cores, and since the emissivity scales as the density squared, there is a boost given to their X-ray luminosities leading to a higher normalisation in the $L_{\mathrm{X}}-T$ relation compared to other sub-samples. The steepening could be explained as a higher increase in X-ray luminosity for relatively high temperature SCC groups ($\geq~1$ keV) which probably goes on decreasing as we go down the temperature scale. The relatively low X-ray luminosity for the NSCC groups (which are WCC and NCC groups) is an indication that there isn't strong cooling going on in these systems and could also indicate the further influence of AGN feedback (particularly for the NCC groups) which suppresses the X-ray luminosity, though we do wish to point out that not all WCC groups harbour a CRS.

\begin{figure}
 \centering
 \includegraphics[scale=0.35, angle=270]{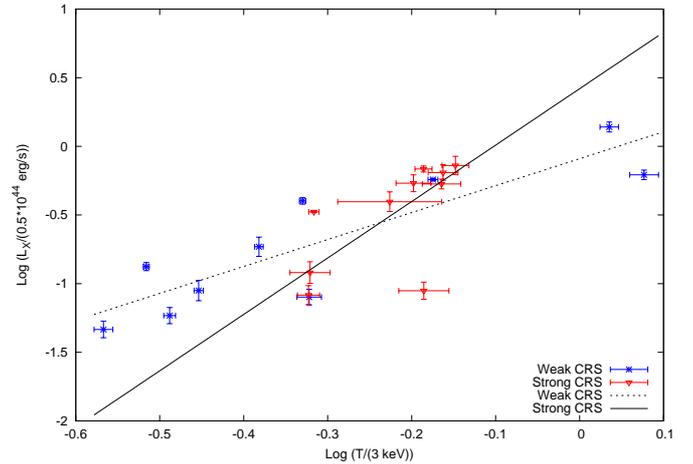}
\caption{The $L_{\mathrm{X}}-T$ relation for sub-samples divided on the basis of the median radio luminosity of the CRS. Blue points (`weak' CRS) represent those groups which have a CRS with total radio luminosity less than the median radio luminosity, while red points represent those groups with a CRS greater than or equal to the median radio luminosity.}\label{GLxTCRSstrength}
 \end{figure}

This brings us to our next point of discussion, namely the $L_{\mathrm{X}}-T$ relation for those groups with and without a CRS. As the slopes and normalisations for the CRS and the NCRS are both consistent within the errors for the CRS and the NCRS cases, we tried to see if divisions based on the morphology of the CRS could unravel some features. Dividing the groups for those with a CRS that show extended radio emission and those with CRSs that show only central emission based on a visual inspection of the radio contours (Appendix C of \citealt{2014arXiv1402.0868B}), we observe that the former has a much steeper slope than the latter ($3.64\pm1.21 $ vs. $2.07\pm0.31$). Taking this one step further, we fit the $L_{\mathrm{X}}-T$ relation for two more sub-samples; this time dividing the sample on the basis of the median radio luminosity of the CRSs (Fig.~\ref{GLxTCRSstrength}). Once again, we observe that the groups with a `strong' radio source (greater than or equal to the median radio luminosity) have a much steeper scaling relation ($4.11\pm1.38$ vs. $1.96\pm0.31$) than those with `weak' radio sources. Qualitatively, both these results are in agreement with \cite{2007MNRAS.379..260M} who find a much steeper slope for objects with extended radio sources ($\sim 4$) than for radio sources with point-like emission. Interestingly, the mean radio output of the SCC groups CRS is a factor of 34 lower than the NSCC groups CRS, which could indicate that AGN activity is less strong in SCC groups than the NSCC groups, assuming the radio luminosity is a good indicator of AGN activity. This would be very much in line with simulation results by \cite{2011MNRAS.415.1549G,2014ApJ...783L..10G} who argue that AGN feedback in galaxy groups must be self-regulated with low mechanical efficiencies, and is a fundamental requirement for the preservation of the cool-core. Moreover, in this particular group sample, 60\% of the groups with a `strong' CRS are NSCC groups while 60\% of the groups with a `weak' CRS are SCC groups. This could be an indication that the more powerful radio AGN are preferentially located in NSCC groups rather than SCC groups. These conclusions are subject to selection effects, and more robust results would require a much larger, complete group sample, backed by radio data down to lower frequencies and more homogeneous flux limits.

\section{Summary}
With a sample of 26 galaxy groups, we studied the effects of ICM cooling, AGN feedback and selection effects on the $L_{\mathrm{X}}-T$ relation on the galaxy group regime. The main results of this study can be summarised as follows:
\begin{itemize}
\item The observed $L_{\mathrm{X}}-T$ relation for groups is significantly affected by selection effects which impact the slope, normalisation and the intrinsic scatter in $L_{\mathrm{X}}$ and requires bias corrections. We also conclude that SCC groups require a different bias correction than NSCC groups.
\item The bias-corrected slope of the $L_{\mathrm{X}}-T$ relation for groups obtained after correcting for selection effects shows indications of steepening, but is consistent within errors to the bias-corrected slope of the clusters. 
\item The SCC groups have the highest normalisation and the steepest slope for the scaling relation. This is attributed to the enhanced luminosities of these systems. 
\item The bias corrected intrinsic scatter in $L_{\mathrm{X}}$ seems to generally go up as we enter the group regime. Additionally, for groups, the observed intrinsic scatter is lower than the bias corrected scatter obtained from simulations for most cases.
\item Subject to selection effects, we see indications of a steepening of the scaling relation for those groups which have a CRS radio luminosity greater than the median, and speculate that such relatively powerful CRSs are preferentially located in NSCC groups.
\end{itemize}
In short, we have demonstrated that the behaviour of the $L_{\mathrm{X}}-T$ relation in groups is similar (e.g. the slopes) and yet different (e.g. intrinsic scatter in $L_{\mathrm{X}}$) to that in clusters. The next step to have a survey ready $L_{\mathrm{X}}-T$ relation would be to account for every process at work in galaxy groups, their effect on the slopes and normalisations, and their contribution to the overall scatter. Quantifying this will be a challenge and as we have pointed out, objectively selected, large samples of groups, particularly to much lower temperatures ($< 1$ keV) with good quality X-ray data, without potential archival bias, would be paramount. Additionally, we expect more features to be obtained when we sub-classify these potential group samples into SCC, WCC and NCC classes and sub-classes thereof, unlike what is done in this study. This is a project that we hope to pursue in the near future.

\begin{acknowledgements}
{V.~B. would like to thank Gerrit Schellenberger for useful discussions and for providing a C version of the BCES code. V.~B. acknowledges support from grant RE 1462/6.~L.~L. acknowledges support by the DFG through grant LO2009/1-1, RE 1462/6 and by the Transregional Collaborative Research Centre TRR33 ``The Dark Universe" (project B18). T.~H.~R. acknowledges support from the DFG through the Heisenberg research grant RE 1462/5.}
\end{acknowledgements}

\bibliographystyle{aa}
\bibliography{ref}
\end{document}